\documentclass[prd,twocolumn,showpacs,superscriptaddress,nofootinbib]{revtex4}

\usepackage{amsfonts}
\usepackage{amsmath}
\usepackage{amssymb}
\usepackage{amsthm}
\usepackage{bm}
\usepackage{dcolumn}
\usepackage{epsfig}
\usepackage{graphicx}
\usepackage{graphics}
\usepackage[latin1]{inputenc}
\usepackage{latexsym}
\usepackage{rotating}
\usepackage{hyperref}

\newcommand\be{\begin{equation}}
\newcommand\ba{\begin{eqnarray}}
\newcommand\ee{\end{equation}}
\newcommand\ea{\end{eqnarray}}

\newcommand{\pont}{{\,^\ast\!}R\,R}

\newcommand{\Scri}{{\mathcal I}}

\begin{document}
\title{Dynamical Chern-Simons Modified Gravity I: \\ Spinning Black Holes in the Slow-Rotation Approximation}

\author{Nicol\'as Yunes}
\affiliation{Department of Physics, Princeton University, Princeton, NJ 08544, USA.}

\author{Frans Pretorius}
\affiliation{Department of Physics, Princeton University, Princeton, NJ 08544, USA.}

\date{\today}

\begin{abstract}

The low-energy limit of string theory contains an anomaly-canceling correction to the Einstein-Hilbert action, which defines an effective theory: Chern-Simons (CS) modified gravity. The CS correction consists of the product of a scalar field with the Pontryagin density, where the former can be treated as a background field (non-dynamical formulation) or as an evolving field (dynamical formulation). Many solutions of general relativity persist in the modified theory; a notable exception is the Kerr metric, which has sparked a search for rotating black hole solutions. Here, for the first time, we find a solution
describing a rotating black hole within the dynamical framework, and in the small-coupling/slow-rotation limit. The solution is axisymmetric and stationary,
constituting a deformation of the Kerr metric with dipole scalar ``hair,'' whose effect on geodesic motion is to weaken the frame-dragging effect and shift the location of the inner-most stable circular orbit outwards (inwards) relative to Kerr for co-rotating (counter-rotating) geodesics. We further show that the correction to the metric scales inversely with the fourth power of the radial distance to the black hole, suggesting it will escape any meaningful bounds from weak-field experiments. For example, using binary pulsar data we can only place an initial bound on the magnitude of the dynamical coupling constant of $\xi^{1/4} \lesssim 10^{4} \; {\textrm{km}}$. More stringent bounds will require observations of inherently strong-field phenomena.

\end{abstract}





\maketitle

\section{Introduction}
\label{intro}

Supermassive, spinning black holes (BHs) are expected to be present at the center 
of most galaxies. The gravitational field in the exterior of such bodies plays a critical
role in the evolution of captured compact objects and in the emission of gravitational
waves (GWs). In General Relativity (GR), such a field is described by the Kerr metric~\cite{Kerr:1963ud}, 
which is a stationary and axisymmetric solution, parameterized exclusively in terms of the 
mass of the BH and its spin angular momentum.  In modified theories of gravity, the Kerr metric need not be a solution to the field equations. 
For example, in Einstein-Dilaton-Gauss-Bonnet gravity, slowly rotating BH solutions were recently
found that differ from Kerr~\cite{Pani:2009wy}.
A measured deviation from the Kerr metric, either from electromagnetic observations~\cite{Psaltis:2007cw,Psaltis:2008bb} or purely gravitational wave ones~\cite{Collins:2004ex,Glampedakis:2005cf}, can therefore provide insight into extensions of GR, or lack thereof.  

One particular theory that has received remarkable attention in recent years
is Chern-Simons (CS) modified gravity~\cite{Jackiw:2003pm}, which also does
not have the Kerr metric as a solution~\cite{Campbell:1990fu}. The action for CS modified gravity is defined by the sum of the 
Einstein-Hilbert action and a new parity-violating, four-dimensional 
correction. Interest in the model spiked when it was found that string theory unavoidingly requires
such a correction to remain mathematically consistent~\cite{Alexander:2004us,Alexander:2004xd}. 
In the perturbative string sector, such a correction is necessary by the Green-Schwarz anomaly-canceling mechanism 
upon four-dimensional compactification (see eg.~\cite{Campbell:1992hc,Alexander:2004xd}. In general, such a correction arises in the presence of Ramond-Ramond scalars due to duality symmetries~\cite{Polchinski:1998rr}. 

The CS correction to the action consists of the product of a {\it{CS scalar field}}, $\vartheta$, and the {\it{Pontryagin density}} $\pont$. The latter is defined as the contraction of the Riemann tensor with its dual. The dual to the Riemann tensor contains contractions of the Levi-Civita tensor, which is odd under a parity transformation, thus potentially enhancing gravitational parity-breaking. Of course, this fact does not imply that parity-preserving solutions are disallowed in CS modified gravity, since as $\vartheta \to {\textrm{const.}}$ the modified theory reduces to GR. Instead, the CS correction introduces a means to enhance parity-violation through a pure curvature term, as opposed to through the matter sector, as more commonly happens in GR.  

Two formulations of CS modified gravity exist that are actually independent theories on their own right, leading to different observables. In the {\it{dynamical formulation}}, the CS scalar is treated as a dynamical field, endowed with its own stress-energy tensor and evolution equation, while in the {\it{non-dynamical formulation}}, the CS scalar is an {\it{a priori}} prescribed function, and its effective evolution equation reduces to a differential constraint on the space of allowed solutions (the so-called Pontryagin constraint, defined as the vanishing of the Pontryagin density). 

Most studies to-date have concentrated on the non-dynamical formulation of CS modified gravity. In particular, the CS modification
has been used to propose an explanation to the leptogenesis problem~\cite{alexander:2004:lfg}, as well as to explain 
the flat rotation curves of galaxies~\cite{Konno:2008np}. A short list of such studies includes some on fermion interactions~\cite{Alexander:2008wi}, 
torsion~\cite{Cantcheff:2008qn,Alexander:2008wi}, the far-field behavior~\cite{Alexander:2007vt,Alexander:2007zg,Smith:2007jm,Yunes:2008ua}, 
GWs~\cite{Alexander:2007:gwp,Yunes:2008bu}, the slow-rotation limit~\cite{Konno:2007ze}, exact solutions~\cite{Tekin:2007rn,Guarrera:2007tu,Grumiller:2007rv} and Schwarzschild BH perturbation theory~\cite{Yunes:2007ss}. In contrast,
the dynamical version of CS modified gravity remains completely unexplored. 

A common thread in the above results concerns the well-posedness and arbitrariness of the non-dynamical formulation. Indeed, in the non-dynamical framework, there are no well-motivated physical reasons 
for particular choices of the scalar field, which is usually chosen {\it{ad-hoc}}
with the goal of simplifying the equations. 
Furthermore, for a given choice of 
CS scalar the Pontryagin constraint significantly restricts the class of allowed
solutions, even to the point where the non-dynamical theory may be over-constrained
and lack a well-posed initial value problem~\cite{Yunes:2007ss}.
In the dynamical framework the arbitrariness is reduced or even completely eliminated, as though one can freely prescribe the initial conditions for the field, the wave-like equation
it must subsequently satisfy could conceivably force the field to evolve to
a unique, late time solution independent of the initial conditions for a large class of spacetimes.
Whether the dynamical framework offers a well-posed initial value problem is presently
unknown.

Only partial results exist on rotating BH solutions in CS modified gravity, which we classify  in Table~\ref{status-matrix}~\footnote{Henceforth, we use the term ``BH solution'' lightly, since all these investigations are in the perturbative regime, where one cannot formally speak of an event horizon, which is the defining feature of a BH. }. 
The capitalized letters that appear on this table stand for the authors of the papers that dealt with the respective investigations, 
where a boldface font means that astrophysically plausible solutions were found.  
As one can see in the table, no work has been done in the dynamical formulation (except for  
some ending remarks in~\cite{Grumiller:2007rv}). In the non-dynamical framework, solutions can be divided into 
two groups: axisymmetric and non-axisymmetric solutions. The second group is allowed in non-dynamical CS modified gravity because 
the gradient of the CS scalar selects a spacetime direction that need not be co-aligned with the BH spin, thus breaking
axisymmetry. 
\begin{table}[]
\begin{tabular}{| l | c | c |}
	\hline
 &  Non-dynamical &  Dynamical \\
	\hline
Axisymmetric  &  {\bf{KMT}}~\cite{Konno:2007ze}, GY~\cite{Grumiller:2007rv}
& -- \\ \hline
Non-Axisymmetric  &  {\bf{AY}}~\cite{Alexander:2007zg,Alexander:2007vt}
& -- \\  \hline
\end{tabular}
\caption{Status of investigations of rotating BH solution in CS modified gravity.}
\label{status-matrix}
\end{table}

The first rotating BH solution was found by Alexander and Yunes (AY)~\cite{Alexander:2007vt,Alexander:2007zg} using a {\it{far-field approximation}} 
(where field point distance is assumed much larger than the BH mass) and $\vartheta$ linearly proportional to the asymptotic time coordinate $t$ (the so-called canonical choice). 
This far-field solution is stationary though not axisymmetric, leading to a correction to the frame-dragging effect, which was later used to 
constrain the theory~\cite{Smith:2007jm,Yunes:2008ua}. The second rotating BH solution was found by Konno, et.~al.~(KMT)~\cite{Konno:2007ze},
using a {\it{slow-rotation approximation}}~\cite{Thorne:1984mz,Hartle:1968si} (where the spin angular momentum is assumed much smaller than the BH mass)
and considering both the canonical choice of $\vartheta$ as well as a new (KMT) choice. KMT showed that a slowly-rotating solution cannot exist with a canonical $\vartheta$, though with the KMT choice they found a stationary and axisymmetric solution with a CS correction that also leads to modified frame-dragging. 

The only study that has searched for exact, rotating BH solutions is that of Grumiller and Yunes (GY)~\cite{Grumiller:2007rv}, who focused on stationary and axisymmetric line elements for arbitrary $\vartheta$. GY argued against the existence of such highly symmetric BH solutions for arbitrarily large spins in the non-dynamical framework, because, although the modified field equations can be satisfied to linear order in the spin, the Pontryagin constraint generically cannot be satisfied to higher order.

Both the far-field solution of AY and the slowly-rotating solution of KMT served greatly to understand the effect of the CS correction on physical observables, 
however these solutions are rather special. First, they have been developed purely with the non-dynamical framework in mind, which as mentioned suffers from potential well-posedness problems in addition to offering no physical or dynamical reasons for the particular
choice of $\vartheta$. Second, these solutions do not carry over to the dynamical
theory, as their choices for $\vartheta$ posses infinite energy, and thus do not
describe a self-consistent perturbation of the GR solutions. These issues are discussed further in Appendix~\ref{rot-BH-non-dym}, where 
we extend the KMT result and explicitly show that it is non-unique, even for the KMT choice of CS scalar.

One is then led to reconsider the question of what metric represents a spinning BH in CS modified gravity with a well-defined CS coupling field that can be embedded in the dynamical framework. That is the topic of this paper, 
the first in a series attempting to shed light on the nature of dynamical CS modified gravity. Here we restrict attention to stationary, axisymmetric perturbations of the slow-rotation
limit of the Kerr metric. In subsequent work we will relax these assumptions, which will most likely
require numerical solution methods.

To summarize the main results, the metric and scalar field describing
the leading order corrections to the Kerr metric in dynamical CS modified gravity are:
\ba
\label{hairy-sol}
ds^{2} &=& ds^{2}_{\textrm{K}}  +
\frac{5}{4} \frac{\alpha^{2}}{\beta \kappa} \frac{a}{r^{4}}\left( 1 + \frac{12}{7} \frac{M}{r} + \frac{27}{10} \frac{M^{2}}{r^{2}} \right) \sin^{2}{\theta} dt d\phi,
\nonumber \\
\vartheta &=&  \frac{5}{8} \frac{\alpha}{\beta} \frac{a}{M} \frac{\cos(\theta)}{r^2} \left(1 + \frac{2 M}{r} + \frac{18 M^2}{5 r^2} \right),
\ea
where $ds^{2}_{\textrm{K}}$ is the slow-rotation limit of the Kerr metric, and
$\alpha$ and $\beta$ are coupling constants in the CS correction to the action (defined in Sec.~\ref{ABC}),
$M$ is the BH mass and $a\cdot M$ is the BH angular momentum. 
Equation~\eqref{hairy-sol} constitutes the first rotating BH solution in dynamical CS modified
gravity, and can be thought of as a small deformation of a Kerr BH with the addition
of CS scalar ``hair''. 
Given that this correction is in the $\phi-t$ sector of the metric, there
is a modification to the frame-dragging effect, though it is suppressed by three powers of the inverse radius relative to the GR solution, implying that weak-field experiments are incapable of placing interesting bounds on the theory. In analogy to the post-Newtonian approximation~\cite{Blanchet:2002av}, the CS correction would correspond to a term of third post-Newtonian order relative to the Kerr term.
To this order, the location of the event horizon and ergosphere of the BH
are unchanged.

An outline of the rest of the paper follows.
In Sec.~\ref{ABC} we review the basics of CS modified gravity.
In Sec.~\ref{axisym} we solve the dynamical theory for a line element and 
CS coupling field describing slowly rotating BHs, valid for small
CS coupling constants. In Sec.~\ref{properties} we describe some properties of
the new solution, discuss related astrophysical implications, and place
bounds on the strength of the CS coupling constants using binary pulsar data.
Sec.~\ref{conclusions} concludes and discusses possible future research.

For completeness, in Appendix~\ref{rot-BH-non-dym} we discuss aspects of approximate
BHs in the non-dynamical theory: we describe solutions
previously found, present several new solutions, and show that none of these
solutions (new or old) carry over to the dynamical theory.
We have also made an attempt to solve the CS modified equations to all orders 
in the rotation parameter with little success
due to the complicated nature of the equations---this effort is discussed
in Appendix~\ref{rot-fast}.

We use the following conventions in this paper: we work exclusively in
four spacetime dimensions with signature
$(-,+,+,+)$~\cite{Misner:1973cw}, with Latin letters $(a,b,\ldots,h)$
ranging over all spacetime indices, round and
square brackets around indices denote symmetrization and
anti-symmetrization respectively, namely $T_{(ab)}=\frac12
(T_{ab}+T_{ba})$ and $T_{[ab]}=\frac12 (T_{ab}-T_{ba})$, partial
derivatives are sometimes denoted by commas (e.g.~$\partial
\theta/\partial r=\partial_r\theta=\theta_{,r}$), the notation $A_{(m,n)}$ stands for a term of 
${\cal{O}}(m,n)$, which itself stands for terms of ${\cal{O}}(\epsilon^{m})$ or ${\cal{O}}(\zeta^{n})$.
The Einstein summation convention is employed unless otherwise specified, and we
use geometrized units where $G=c=1$.

\section{CS modified gravity}
\label{ABC}
  
In this section, we describe the relevant aspects that define a complete formulation of CS modified gravity and establish some notation 
(for a detailed review see eg.~\cite{Review}). In the first subsection, we present a complete description of the CS modified action and the 
modified field equations of the theory. In the second subsection, we classify the modified theory into two formulations: dynamical versus 
non-dynamical CS modified gravity.
 
\subsection{Basics}

Consider the action
\be
\label{CSaction}
S = S_{\rm EH} + S_{\rm CS} +  S_{\vartheta} + S_{\rm mat},
\ee
where 
\ba
\label{EH-action}
S_{\rm{EH}} &=& \kappa \int_{{\cal{V}}} d^4x  \sqrt{-g}  R, 
\\
\label{CS-action}
S_{\rm{CS}} &=& \frac{\alpha}{4} \int_{{\cal{V}}} d^4x  \sqrt{-g} \; 
\vartheta \; \pont\,,
\\
\label{Theta-action}
S_{\vartheta} &=& - \frac{\beta }{2} \int_{{\cal{V}}} d^{4}x \sqrt{-g} \left[ g^{a b}
\left(\nabla_{a} \vartheta\right) \left(\nabla_{b} \vartheta\right) + 2 V(\vartheta) \right], \quad
\\
S_{\textrm{mat}} &=& \int_{{\cal{V}}} d^{4}x \sqrt{-g} {\cal{L}}_{\textrm{mat}}.
\ea
The first term in Eq.~\eqref{CSaction} is the standard Einstein-Hilbert term, while
the second  one is the CS correction, the third one is the scalar-field term and 
the last one describes additional matter sources, with ${\cal{L}}_{\textrm{mat}}$ 
some matter Lagrangian density. We here employ the 
following conventions: $\kappa^{-1} = 16 \pi G$; $\alpha$ and $\beta$ are {\it{dimensional}} coupling constants,
 $g$ is the determinant of the metric, $\nabla_{a}$ is the covariant derivative associated with the metric tensor $g_{ab}$, 
 and $R$ is the Ricci scalar. The quantity $\pont$ is the Pontryagin density, defined via
\be
\label{pontryagindef}
\pont= R \tilde R = {\,^\ast\!}R^a{}_b{}^{cd} R^b{}_{acd}\,,
\ee
where the dual Riemann-tensor is defined as 
\be
\label{Rdual}
{^\ast}R^a{}_b{}^{cd}=\frac12 \epsilon^{cdef}R^a{}_{bef}\,,
\ee
with $\epsilon^{cdef}$ the 4-dimensional Levi-Civita tensor. 

The {\it{CS coupling field}}, $\vartheta$, is a function of spacetime that parameterizes deformations from GR. If $\vartheta = \textrm{const.}$~CS modified gravity reduces identically to GR because the Pontryagin density is the total divergence of the CS topological current $K^{a}$
\be
\nabla_a K^a = \frac{1}{2} \pont, 
\label{eq:curr1}
\ee
where  
\be
K^a =\epsilon^{abcd} \Gamma^n_{bm} \left(\partial_c\Gamma^m_{dn}+\frac{2}{3} \Gamma^m_{cl}\Gamma^l_{dn}\right)\,,
\label{eq:curr2}   
\ee
and $\Gamma$ is the Christoffel connection. We can use this fact to rewrite $S_{\textrm{CS}}$ in a more standard form via integration by parts:
\be
\label{CS-action-K}
S_{\textrm{CS}} =  \alpha 
\left( \vartheta \; K^{a} \right)|_{\partial {\cal{V}}}
-
 \frac{\alpha}{2} \int_{{\cal{V}}} d^4x  \sqrt{-g} \; 
\left(\nabla_{a} \vartheta \right) K^{a}.
\ee
The first term is usually discarded since it is evaluated on the boundary of the manifold~\cite{Grumiller:2008ie}, while the second term corresponds to the CS correction. 

The modified field equations are obtained by variation of the action with respect to the metric and the CS coupling field:
\ba
\label{eom}
R_{ab} + \frac{\alpha}{\kappa} C_{ab} &=& \frac{1}{2 \kappa} \left(T_{ab} - \frac{1}{2} g_{ab} T \right),
\\
\label{eq:constraint}
\beta \; \square \vartheta &=& \beta \; \frac{dV}{d\vartheta} - \frac{\alpha}{4} \pont,
\ea
where $R_{ab}$ is the Ricci tensor and $\square = \nabla_{a} \nabla^{a}$ is the D'Alembertian operator.
The quantity $C_{ab}$ is the so-called C-tensor
\be
\label{Ctensor}
C^{ab} = v_c
\epsilon^{cde(a}\nabla_eR^{b)}{}_d+v_{cd}{\,^\ast\!}R^{d(ab)c}\,,   
\ee
where 
\be
\label{v}
v_a=\nabla_a\vartheta\,,\qquad
v_{ab}=\nabla_a\nabla_b\vartheta.
\ee 
The total stress-energy tensor is
\be\label{Tab-total}
T_{ab} = T^{\textrm{mat}}_{ab} + T_{ab}^{\vartheta}, 
\ee
where $T^{\textrm{mat}}_{ab}$ is the contribution from
other matter sources (which we will set to zero here),
and $T_{ab}^{\vartheta}$ is the scalar field contribution,
given by
\be
\label{Tab-theta}
T_{ab}^{\vartheta} 
=   \beta  \left[  \left(\nabla_{a} \vartheta\right) \left(\nabla_{b} \vartheta\right) 
    - \frac{1}{2}  g_{a b}\left(\nabla_{a} \vartheta\right) \left(\nabla^{a} \vartheta\right) 
-  g_{ab}  V(\vartheta)  \right].
\ee

The strong equivalence principle ($\nabla_{a} T^{ab}_{\textrm{mat}} = 0$) is naturally preserved in CS modified gravity, provided the equations of motion for $\vartheta$ hold [Eq.~\eqref{eq:constraint}]. This is because when one takes the divergence of Eq.~\eqref{eom}, the first term on the left-hand side vanishes by the Bianchi identities, while the second term is proportional to the Pontryagin density via
\be
\label{nablaC}
\nabla_a C^{ab} = - \frac{1}{8} v^b \pont.
\ee
The equality of this contribution to the divergence of the scalar field stress-energy tensor leads to Eq.~\eqref{eq:constraint}.

\subsection{Two formulations of CS modified gravity}
\label{dyn-vs-can}

The modified theory can be classified into two distinct formulations: dynamical and non-dynamical.  
The non-dynamical formulation is defined by setting $\beta = 0$~\footnote{Traditionally, $\alpha = \kappa$ when working in this formulation, but this is actually not necessary and we choose to leave this constant arbitrary.}, in which case the field equations become 
\ba
\label{eom-nd}
R_{ab} + \frac{\alpha}{\kappa} C_{ab} &=& \frac{1}{2 \kappa} \left(T_{ab}^{\textrm{mat}}- \frac{1}{2} g_{ab} T^{\textrm{mat}} \right),
\nonumber \\
\label{eq:constraint-nd}
0 &=& \pont.
\ea
In vacuum, the right-hand side of the first equation identically vanishes. The second equation, the Pontryagin constraint, which used to be an evolution equation for $\vartheta$, has now become an additional differential constraint on the space of allowed solutions.

In the non-dynamical framework not only does the Pontryagin constraint reduce 
the space of allowed solutions, one must also prescribe {\it{a priori}} 
the entire history of the CS coupling $\vartheta$. Once this prescription is made, the CS coupling field is effectively frozen and unaffected by any interaction. The so-called canonical choice of $\vartheta$ is 
\be
\vartheta_{\textrm{can}} = \frac{t}{\mu},
\qquad
v^{a}_{\textrm{can}} = \left[\frac{1}{\mu},0,0,0\right],
\ee
which was chosen when the non-dynamical framework was first postulated~\cite{Jackiw:2003pm}. In hindsight, there is nothing truly ``canonical'' about this choice of CS scalar, except that it simplifies the modified field equations dramatically. Moreover, it is clearly not gauge-invariant, and the theory offers no motivation or argument
why some particular slicing of spacetime is preferred.
The canonical choice has been found to be very restrictive, not allowing for axisymmetric rotating BH solutions~\cite{Alexander:2007zg,Alexander:2007vt,Konno:2007ze} and perturbations of a certain parity of Schwarzschild BHs~\cite{Yunes:2007ss}.  

The dynamical formulation allows $\beta$ to be arbitrary, in which case the modified field equations are given by 
Eqs.~\eqref{eom}-\eqref{Tab-theta}.
Equation~\eqref{eq:constraint} lifts the Pontryagin constraint, and is 
now an evolution equation for the CS coupling field. Therefore, no restriction is imposed {\it{a priori}} on the allowed space of solutions. Instead of prescribing the entire history of the CS coupling, one specifies some initial conditions for $\vartheta$, which then evolves self-consistently via Eq.~\eqref{eq:constraint}.

The dynamical and non-dynamical formulations are inequivalent and independent theories, 
despite sharing similarities in the action and a few of the same solutions.  
Although one can take the limit $\beta \to 0$ at the level of the action to obtain the non-dynamical
framework, one cannot expect that the same limit acting on {\emph{solutions}} of the dynamical theory 
would return solutions of the non-dynamical framework. An intuitive way to see this is to consider ever
smaller $\beta$ parameters in the scalar field evolution equation. Since generically the Pontryagin
density is non-vanishing, an ever smaller $\beta$ forces an immense scalar field, whose bare value 
then couples to the metric via the C-tensor, leading to an equally large back-reaction on the geometry.    
Therefore, as one ``freezes'' the scalar field in the dynamical theory, one will generically violate its evolution
equation, leading to divergences as exemplified by the solution we found in Eq.~\eqref{hairy-sol}.

We conclude this section with a few words on the dimensions of the coupling
constants and scalar field. The choice of units of one of
$(\alpha,\beta,\vartheta)$ constrains the units for the others.
For example, if we wish the CS scalar field to have units $[\vartheta] = L^{A}$, then $[\alpha] = L^{2 - A}$ and $[\beta] = L^{-2A}$, where $L$ is a unit of length and where we have set $[\kappa] = 1$. A more natural choice may be to require the CS scalar to be dimensionless, as is done in scalar tensor theories, which would then require that $[\alpha] = L^{2}$ and $\beta$ be dimensionless~\footnote{Note here that we are employing geometric units where $G = c = 1$, and thus, the action has units of $L^{2}$. Had we employed natural units $h = c = 1$, then the action would be dimensionless and if $[\vartheta] = L^{A}$ then $[\alpha] = L^{-A}$ and $[\beta] = L^{-2 A - 2}$.}. Another natural choice would be to set $\beta = \alpha$, thus putting $S_{\vartheta}$ and $S_{CS}$ on the same footing; we would then have
$[\vartheta] = L^{-2}$. Neither formulation requires that we choose specific units for $\vartheta$, so we shall leave these arbitrary, as results in the literature have made a variety of different choices. 

\section{Rotating Black Holes in Dynamical CS Modified Gravity }\label{axisym}

In this section we study rotating BHs in the dynamical formulation of the modified theory. 
The analytical study of stationary and axisymmetric line elements in this theory
without the aid of any approximation scheme is a quixotic task. We therefore employ
a couple of approximations, and begin by defining and explaining them. We then
proceed by to solve the modified field equations to second order in the
perturbative expansion.

\subsection{Approximation Scheme}
\label{approx}

We choose to employ two approximations schemes: a {\it{small-coupling}} approximation and a {\it{slow-rotation}} approximation.
In the small-coupling scheme we shall treat the CS modification as a small deformation of GR, which allows us to pose the following metric decomposition (to second order):
\be
g_{ab} = g_{ab}^{(0)} + \zeta g^{(1)}_{ab}(\vartheta) + \zeta^{2} g^{(2)}_{ab}(\vartheta),
\label{small-cou-exp0}
\ee
where $g_{ab}^{(0)}$ is some background metric that satisfies the Einstein equations, such as the Kerr metric, while $g_{ab}^{(1)}(\vartheta)$ and $g_{ab}^{(2)}(\vartheta)$ are first and second-order CS perturbations that depend on $\vartheta$. The book-keeping parameter $\zeta$ labels the order of the small-coupling approximation.

The slow-rotation scheme allows us to re-expand both the background and the $\zeta$-perturbations in powers of the Kerr rotation parameter $a$. The background metric and the metric perturbation then become
\ba
\label{small-cou-exp}
g_{ab}^{(0)} &=& \eta_{ab}^{(0,0)} + \epsilon \; h_{ab}^{(1,0)} + \epsilon^{2} h_{ab}^{(2,0)},
\nonumber \\
\zeta g_{ab}^{(1)} &=& \zeta h_{ab}^{(0,1)} + \zeta \epsilon \; h_{ab}^{(1,1)} + \zeta \epsilon^{2} h_{ab}^{(2,1)},
\nonumber \\
\zeta^{2} g_{ab}^{(2)} &=& \zeta^{2} h_{ab}^{(0,2)} + \zeta^{2} \epsilon \; h_{ab}^{(1,2)} + \zeta^{2} \epsilon^{2} h_{ab}^{(2,2)},
\ea
where the book-keeping parameter $\epsilon$ labels the order of the slow-rotation expansion. We should recall here that the notation $h^{(m,n)}_{ab}$ stands for terms of ${\cal{O}}(m,n)$, which in turn stands for a term of ${\cal{O}}(\epsilon^{m})$ and ${\cal{O}}(\zeta^{n})$. For example, in Eq.~\eqref{small-cou-exp}, $\eta_{ab}^{(0,0)}$ is the background metric in the limit $a = 0$, while $h_{ab}^{(1,0)}$ and $h_{ab}^{(2,0)}$ are first and second-order expansions of the background metric in the spin parameter. 

Combining both approximation schemes we obtain a bivariate expansion in two independent parameter $\zeta$ and $\epsilon$, which to second order is given by
\be
g_{ab} = \eta_{ab}^{(0,0)} + \epsilon h_{ab}^{(1,0)} + \zeta h_{ab}^{(0,1)} + \epsilon \zeta h_{ab}^{(1,1)} + \epsilon^{2} h_{ab}^{(2,0)} + \zeta^{2} h_{ab}^{(0,2)},
\ee
When we refer to first-order terms, we shall mean terms of ${\cal{O}}(1,0)$ or ${\cal{O}}(0,1)$, while when we say second-order terms we refer to those of ${\cal{O}}(2,0)$, or ${\cal{O}}(0,2)$ or ${\cal{O}}(1,1)$. 

What precisely are these book-keeping parameter $\epsilon$ and $\zeta$? The slow-rotation scheme is an expansion in the Kerr parameter, and thus its {\it{dimensionless}} expansion parameter must be $a/M$. Therefore, a term in the equations multiplied by $\epsilon^n$ is of ${\cal{O}}\left((a/M)^n\right)$.
The small-coupling expansion parameter must depend on the ratio of CS coupling to the GR coupling, $\alpha/\kappa$, because such a combination multiplies the C-tensor in Eq.~\eqref{eom}. The definition of the C-tensor [Eq.~\eqref{Ctensor}] clearly states that this tensor is proportional to gradients of the CS scalar, which itself must be proportional to $\alpha/\beta$ due to the $\vartheta$-evolution equation [Eq.~\eqref{eq:constraint}]. We see then that the CS correction to the metric will be proportional to the combination $\xi = (\alpha/\kappa) (\alpha/\beta)$. Such a factor, however, is not dimensionless, and thus, it cannot be formally treated as a perturbation parameter. The only mass scale available is that of the background metric, which to leading order in the slow-rotation expansion is simply the BH mass. We shall then choose to normalize $\xi$ such that the book-keeping parameter is $\zeta$ multiplies terms of ${\cal{O}}\left[\alpha^{2}/(\kappa \beta M^4)\right]$.

The small-coupling scheme together with the structure of the modified field equations establish a well-defined iteration or boot-strapping scheme. From Eq.~\eqref{eom} and~\eqref{eq:constraint}, one can see that the source of the $\vartheta$-evolution equation is always of lower order than the CS correction to the Einstein equations. Such an observation implies that we can independently solve the evolution equation for $\vartheta$ first. The solution obtained for $\vartheta$ can then be used in the modified field equations to find the CS correction to the metric. 
In principle this procedure can then be iterated to find solutions
to higher order in the expansion parameters.

Let us illustrate such a boot-strapping scheme in more detail. To facilitate this, let us temporarily choose units $\beta = \kappa$, such that $\vartheta$ is dimensionless. The small-coupling expansion parameter is then controlled by $\alpha$ only via $\zeta = {\cal{O}}[\alpha^{2}/(\kappa^{2} M^{4})]$. In these units, we see then that the right-hand side of Eq.~\eqref{eq:constraint} is proportional to $\zeta^{1/2}$, while the second term in Eq.~\eqref{eom} is proportional to $\zeta$. In turn, this implies that $\vartheta$ is a Frobenius series with fractional structure 
\be
\vartheta = \zeta^{1/2} \sum_{n=0}^{\infty} \zeta^{n} \vartheta^{(n)},
\ee
while the metric perturbation is a regular series in natural powers of $\zeta$, as required in Eq.~\eqref{small-cou-exp0}. 

Alternatively, one could use different units for $\vartheta$ that slightly change the order counting. For example, let us temporarily choose units $\beta = \alpha$, which by dimensional analysis automatically implies $[\alpha] = L^{4}$ and $[\theta] = L^{-2}$. The small coupling parameter then becomes $\zeta = {\cal{O}}(\alpha/M^{4})$, while the right-hand side of Eq.~\eqref{eq:constraint} is now proportional to $\zeta^{0}$ to leading order. In these units both $\vartheta$ and $g_{ab}$ have expansions in natural powers of $\zeta$, however the leading-order term of the former is $1/\zeta$ larger than the latter. 

Irrespective of units, we see then that the $\vartheta$-evolution equation is always of lower order relative to the modified field equation, which leads to a well-defined boot-strapping scheme.  Whether a term proportional to $(a/M) (\alpha/\beta)$ is of ${\cal{O}}(1,1)$ or ${\cal{O}}(1,1/2)$ depends on the choice of $\beta$. For the purposes of order counting only, we shall assume that $\beta \propto \alpha$, though we shall leave all factors of $\beta$ explicit. With this choice, $\alpha/\beta$ is of order unity and both $\vartheta$ and $g_{ab}$ have expansions in natural powers of $\zeta$. 

This boot-strapping scheme is, in a sense, analogous to the semi-relativistic approximation~\cite{Ruffini:1981rs}, where one models extreme-mass ratio inspirals by solving the geodesic equations and neglecting the self-force of the particle. Even with this approximation, if the background is sufficiently complicated (eg.~the Kerr metric), one will not be able to solve the field equations. The small-rotation scheme is then introduced such that the independent equations, derived from the bootstrapping scheme in the small-coupling approximation, can actually be solved analytically.  

Note that in the bivariate expansions we require that
both $\zeta$ and $\epsilon$ are {\it{independently}} small. We shall not impose any restrictions on their relative size, except for requiring that $\epsilon$ not be proportional to an inverse power of $\zeta$, since this would break the above requirement. We emphasize again that as is commonly done in perturbation theory $\epsilon$ and $\zeta$ are only {\it{bookkeeping}} parameters and are not equal to $a/M$ or $\xi/M^{4}$, rather they multiply terms in the resultant
equations which are of that order.
Since these parameters do not carry any physical meaning, we can and shall set them to unity at the end of the calculation.  

\subsection{Slowly Rotating BH Solutions}\label{slow-rot}

The slow-rotation expansion of the background metric can be formalized via the Hartle-Thorne approximation~\cite{Thorne:1984mz,Hartle:1968si}, where the line element is parameterized by
\ba
\label{slow-rot-ds2}
ds^{2} &=& - f  \left[1 + h(r,\theta)\right] dt^{2}
\nonumber \\
&+& \frac{1}{f}  \left[1 + m(r,\theta)\right] dr^{2}
\nonumber \\
&+& r^{2} \left[1 + k(r,\theta) \right] d\theta^{2} 
\nonumber \\
&+& r^{2} \sin^{2}{\theta} \left[1 + p(r,\theta) \right] \left[ d\phi  - \omega(r,\theta) dt \right]^{2},
\ea
where $M$ is the BH mass in the absence of the CS correction and $f = 1 - 2 M/r$ is the Schwarzschild factor. 
In Eq.~\eqref{slow-rot-ds2}, $(t,r,\theta,\phi)$ are Boyer-Lindquist coordinates and the metric
perturbations are $h(r,\theta)$, $m(r,\theta)$, $k(r,\theta)$, $p(r,\theta)$ and $\omega(r,\theta)$.  

The metric in Eq.~\eqref{slow-rot-ds2} has been written as in~\cite{Thorne:1984mz,Hartle:1968si}, but clearly the metric perturbations must be expanded as a series both in $\zeta$ and $\epsilon$. Keeping terms up to second order we have
\ba
 h(r,\theta) &=& \epsilon \; h_{(1,0)} + \epsilon \;  \zeta \; h_{(1,1)} + \epsilon^{2} \; h_{(2,0)},
 \nonumber \\
  m(r,\theta) &=& \epsilon \; m_{(1,0)} + \epsilon \;  \zeta \; m_{(1,1)} + \epsilon^{2} \; m_{(2,0)},
 \nonumber \\
  k(r,\theta) &=& \epsilon \; k_{(1,0)} + \epsilon \;  \zeta \; k_{(1,1)} + \epsilon^{2} \; k_{(2,0)},
 \nonumber \\
  p(r,\theta) &=& \epsilon \; p_{(1,0)} + \epsilon \;  \zeta \; p_{(1,1)} + \epsilon^{2} \; p_{(2,0)}.
 \nonumber \\
  \omega(r,\theta) &=& \epsilon \; \omega_{(1,0)} + \epsilon \;  \zeta \; \omega_{(1,1)} + \epsilon^{2} \; \omega_{(2,0)}.
 \ea
Note that there are no terms of ${\cal{O}}(0,0)$ since these are already included in the Schwarzschild structure of Eq.~\eqref{slow-rot-ds2}. 
Also, we will {\it assume} that in the limit as $a\rightarrow 0$ we uniquely recover Schwarzschild as the solution,
which implies that all terms of ${\cal{O}}(0,n)$ are zero. 
Thus, the CS correction must be at least linear in the Kerr spin parameter $a$.
From the slow-rotation limit of the Kerr metric in GR we can read-off the metric perturbations proportional to $\zeta^{0}$:
\ba
h_{(1,0)} &=& m_{(1,0)} = k_{(1,0)} = p_{(1,0)} =0,
\nonumber \\
\omega_{(1,0)} &=& \frac{2 M a}{r^{3}},
\ea
to first order and 
\ba
m_{(2,0)} &=& \frac{a^{2}}{r^{2}} \left( \cos^{2}{\theta} - \frac{1}{f} \right),
\quad
k_{(2,0)} = \frac{a^{2}}{r^{2}} \cos^{2}{\theta},
\nonumber \\
p_{(2,0)} &=& \frac{a^{2}}{r^{2}} \left(1 + \frac{2 M}{r} \sin^{2}{\theta} \right),
\quad
\omega_{(2,0)} = 0,
\nonumber \\
h_{(2,0)} &=& \frac{2 a^{2} M}{f r^{3}} \left( \cos^{2}{\theta} + \frac{2 M}{r} \sin^{2}{\theta} \right).
\ea
to second order.

All fields must be expanded in the small-coupling and slow-rotation approximation, including the CS coupling field. To get a flavor of the leading-order behavior
of $\vartheta$ we can return to its evolution equation [Eq.~\eqref{eq:constraint}]. We see from this equation that $\partial^{2} \vartheta \sim (\alpha/\beta) \pont$, where
the Pontryagin density vanishes identically to zeroth order in $a/M$. 
Thus, the leading order behavior of the CS scalar must be $\vartheta \sim (\alpha/\beta) (a/M)$, which is clearly always at least proportional to $\epsilon$.
As discussed in Sec.~\ref{approx}, such a term is either of ${\cal{O}}(1,0)$ ($\beta = \alpha$) or 
${\cal{O}}(1,1/2)$ ($\beta = \kappa$),  depending on the choice of $\beta$; here we take the former view. 
Also, from the assumption that the Schwarzschild metric is the unique solution in the zero-angular
momentum limit, we must have that $\vartheta^{(0,n)} = 0$ for all $n$. The expansion for the CS scalar
thus is
\be
\label{th-ansatz}
\vartheta = \epsilon \; \vartheta^{(1,0)}(r,\theta) + \epsilon \;  \zeta \; \vartheta^{(1,1)}(r,\theta) + \epsilon^{2} \; \vartheta^{(2,0)}(r,\theta).
\ee

Let us now begin to apply the algorithm that we described earlier to solve the modified field equations, by focusing first on the 
evolution equation for the CS scalar. To ${\cal{O}}(1,0)$ the evolution equation becomes
\ba
\label{1st-eq}
f \vartheta^{(1,0)}_{,rr} &+& \frac{2}{r} \vartheta^{(1,0)}_{,r} \left( 1 - \frac{M}{r} \right) + \frac{1}{r^{2}} \vartheta^{(1,0)}_{,\theta\theta} + \frac{\cot{\theta}}{r^{2}} \vartheta^{(1,0)}_{,\theta} 
\nonumber \\
&=& - \frac{72 M^{3}}{r^{7}} \frac{\alpha}{\beta}  \frac{a}{M} \cos{\theta},
\ea
where we have set the potential $V(\vartheta) = 0$. The solution to this partial differential equation is a linear superposition of the homogeneous solution and a particular solution: $\vartheta^{(1,0)} = \vartheta^{(1,0)}_{H} + \vartheta^{(1,0)}_{P}$. 
The homogeneous equation is separable:
\be
\vartheta^{(1,0)}_{H}(r,\theta) = \tilde\vartheta(r) \hat\vartheta(\theta).
\ee
The partial differential equation then becomes a set of ordinary differential equations for $\tilde{\vartheta}$ and $\hat\vartheta$, 
whose solution is 
\ba
\label{Hom-sol-1}
\tilde\vartheta(r) &=& E_{1} F\left[\left[\frac{\tilde\alpha}{2},\frac{\tilde\alpha}{2}\right],\tilde\alpha,\frac{2 M}{r} \right] r^{-\tilde\alpha/2} 
\nonumber \\
&+& E_{2} F\left[\left[\frac{\tilde\beta}{2},\frac{\tilde\beta}{2}\right],\tilde\beta,\frac{2 M}{r} \right] r^{-\tilde\beta/2},
\nonumber \\
\hat\vartheta(\theta) &=& E_{3} P_{-\tilde\alpha/2}(\cos{\theta}) + E_{4} Q_{-\tilde\alpha/2}(\cos{\theta}), 
\ea
where $P(\cdot)$ are Legendre polynomials of the first kind, $Q(\cdot)$ are Legendre polynomials of the second kind, 
$F(\cdot)$ are generalized hypergeometric functions, $E_{i}$ are constants of integration and the coefficients
\be
\label{tilde-alpha}
\tilde{\alpha} = 1 - \sqrt{1 - 4 c_{1}},
\qquad 
\tilde{\beta} = 1 + \sqrt{1 - 4 c_{1}},
\ee
where $c_{1}$ is the constant of integration that arises through separation of variables. 

Information about what reasonable constants of integration are can be found by studying the solution $\vartheta^{(1,0)}$ in more detail. Let us first consider the far-field behavior of 
the solution, $r \gg M$, in which limit 
\be
\tilde{\vartheta}(r) \sim E_{1} \left[ 1 + \frac{M}{2 r} \tilde{\alpha} \right] r^{-\tilde{\alpha}/2} 
+  E_{2} \left[ 1 + \frac{M}{2 r} \tilde{\beta} \right] r^{-\tilde{\beta}/2} .
\ee
By requiring that $\vartheta$ be real, we immediately see that $\tilde{\alpha} \in \Re$ and $\tilde{\beta} \in \Re$, which implies $c_{1} < 1/4$. Moreover, if we wish $\vartheta$ to have finite total energy outside of the horizon,
then $\vartheta$ must decay to a constant asymptotically faster than $1/r$,
which implies that $\tilde{\alpha} > 2$ and $\tilde{\beta}>2$. The first requirement cannot be realized for any real $c_{1} < 1/4$, thus forcing $E_{1} = 0$, while the second requirement leads to $c_{1} < 0$. Requiring finite total energy also implies
$\vartheta$ cannot be proportional to $\ln(f)$.
All these considerations then force us to $\vartheta^{(1,0)}_{H} = {\textrm{const}}$. 

Now that the homogeneous solution has been found, we can concentrate on the particular one. One finds
\be
\label{theta-sol-SR}
\vartheta^{(1,0)}_{P} =  \frac{5}{8} \frac{\alpha}{\beta} \frac{a}{M} \frac{\cos(\theta)}{r^2} \left(1 + \frac{2 M}{r} + \frac{18 M^2}{5 r^2} \right) + {\textrm{const.}},
\ee
where we can set the additional integration constant to zero because it does not contribute to the modified Einstein equations~\footnote{Interestingly, the behavior of a scalar field in a Kerr background has already been studied when considering axion hair for Kerr~\cite{Campbell:1990ai,Reuter:1991cb} and dyon~\cite{Campbell:1991rz} BHs and cosmological scenarios~\cite{Kaloper:1991rw}, in the context of string theory. The solution found there is identical to the one found here in Eq.~\eqref{theta-sol-SR}.}.

Now that the CS coupling field has been determined, we can search for CS corrections to the metric perturbations. 
Note that the stress-energy tensor for the CS scalar found here enters the modified field equations at ${\cal{O}}(2,1)$, 
and thus it does not contribute to the metric perturbation. The modified Einstein equations decouple into two groups: 
one that forms a closed system of partial differential equations
for $h^{(1,1)}$, $m^{(1,1)}$, $p^{(1,1)}$ and $k^{(1,1)}$, which consists of the $(t,t)$, $(r,r)$, $(r,\theta)$, $(\theta,\theta)$ and $(\phi,\phi)$-components
of the modified Einstein equations, and another group that consists of a single differential equation for $w^{(1,1)}$, namely the $(t,\phi)$-component
of the modified Einstein equations. 

The first group is independent of the CS coupling field, $\vartheta$, since it arises exclusively from the Ricci tensor. One can verify that with Eq.~\eqref{theta-sol-SR}, 
the relevant components of the C-tensor for the first group vanish exactly. Since these metric perturbations do not constitute a CS deformation (ie.~they are $\zeta$ independent), we can set them to zero: $h^{(1,1)}=0$, $m^{(1,1)}=0$, $k^{(1,1)}=0$ and $p^{(1,1)}=0$. 

The only remaining equation is that from the second group:
\ba
&& 2  \sin^{2}{\theta}  w^{(1,1)}_{,\theta\theta} + 3 \sin{2 \theta} \; w^{(1,1)}_{,\theta} + 8 r f \sin^{2}{\theta} w^{(1,1)}_{,r} 
\nonumber \\
&+& 2 r^2 f \sin^{2}{\theta} w^{(1,1)}_{,rr} = \frac{15}{2} \frac{\alpha^{2}}{\beta \kappa} \frac{a \; f}{r^{8}} \sin^{2}{\theta}
\nonumber \\
&\times& \left(3 r^{2} + 8 M r + 18 M^{2}\right).
\ea
 Once more, the most general solution is a linear combination of a homogeneous solution and a particular
solution. The particular solution is given by 
\be
\label{w-sol-SR}
\omega^{(1,1)} = - \frac{5}{8} \frac{\alpha^{2}}{\beta \kappa} \frac{a}{r^{6}}\left( 1 + \frac{12}{7} \frac{M}{r} + \frac{27}{10} \frac{M^{2}}{r^{2}} \right).
\ee
The homogeneous solution is a sum of generalized hypergeometric functions, whose argument is $r/(2 M)$ and depend on some separation constant $c_{1}$.
For some values of this constant, the solution is purely real but it diverges at least linearly at spatial infinity, while for other values of this constant the solution is either 
complex or infinite.
For this reason, we choose the integration constants that multiply these hypergeometric functions to be zero, thus yielding Eq.~\eqref{w-sol-SR} as the full solution. We note that this perturbation is indeed proportional to $\zeta$ as expected and it has the correct units $[\omega] = L^{-1}$, since $[\xi] = L^{4}$.

The full gravitomagnetic metric perturbation to linear order in $\zeta$ and $\epsilon$ is\footnote{We leave $\kappa$ explicitly here to keep track of relative dimensions between $\alpha$, $\beta$ and $\kappa$ and also to keep track of hidden factors of $16 \pi$.}
\be
\omega = \frac{2 M a}{r^{3}} - \frac{5}{8}  \frac{a \xi }{r^{6}}\left( 1 + \frac{12}{7} \frac{M}{r} + \frac{27}{10} \frac{M^{2}}{r^{2}} \right).
\ee
The above formulae constitute the first slow-rotating BH solution in dynamical CS modified gravity. Note that the perturbation is highly suppressed in the far field limit, decaying as $r^{-6}$, which suggests that its signature can only be observed in the strong field regime. 

As expected, the correction to the metric is a small $\xi$-deformation of the Kerr metric, in agreement with the small-coupling approximation. In fact, one could verify that such an approximation is self-consistent by calculating the next order correction to $\vartheta$. This correction consists of both $\vartheta^{(2,0)}$ and $\vartheta^{(1,1)}$, which can be computed by solving the evolution equation to next order. Doing so, we find
\ba
\vartheta^{(2,0)} &=& 0,
\nonumber \\ \label{theta_11}
\vartheta^{(1,1)} &=& - \frac{25}{448} \frac{\alpha}{\beta} \frac{\xi a}{M^{5}} \frac{\cos{\theta}}{ r^{2}} \left(1 + \frac{2 M}{r} + \frac{18 M^{2}}{5 r^{2}} + \frac{32 M^{3}}{5 r^{3}} 
\right. 
\nonumber \\
&+& \left.
\frac{80 M^{4}}{7 r^{4}} + \frac{144 M^{5}}{7 r^{5}} + \frac{112 M^{6}}{5 r^{6}} + \frac{448 M^{7}}{25 r^{7}} \right),
\ea
which is clearly $\zeta$-times smaller than $\vartheta^{(1,0)}$, thus rendering the small-coupling approximation self-consistent. If we were to use this improved
$\vartheta$ solution in the modified field equation, we would find a correction to the metric proportional to $\zeta^{2} \epsilon$, which we are here neglecting.

\section{Properties of the New Solution}
\label{properties}

We now wish to study some of the geometric properties of the slowly-rotating solution just found, and
comment on astrophysical implications of this.
For completeness, we present the non-vanishing 
metric components below:
\ba\label{sol:metric_elements}
g_{tt} &=& -f  - \frac{2 a^{2} M}{r^{3}} \cos^{2}{\theta}\nonumber, \\
g_{t\phi} &=& - \frac{2 M a}{r} \sin^{2}{\theta} 
\nonumber \\
&+& \frac{5}{8} \frac{\xi}{M^{4}} \frac{a}{M} \frac{M^{5}}{r^{4}} \left(1 + \frac{12 M}{7 r} + \frac{27 M^{2}}{10 r^{2}} \right) \sin^{2}{\theta},
\nonumber \\
g_{rr} &=& \frac{1}{f} + \frac{a^{2}}{f r^{2}} \left(\cos^{2}{\theta} - \frac{1}{f} \right),
\nonumber \\
g_{\theta \theta} &=& r^{2} + a^{2} \cos^{2}{\theta},
\nonumber \\
g_{\phi \phi} &=& r^{2} \sin^{2}{\theta} + a^{2} \sin^{2}{\theta} \left(1 + \frac{2 M}{r} \sin^{2}{\theta} \right),
\ea
and recall that this metric is accurate to ${\cal{O}}(2,0)$, ${\cal{O}}(1,1)$ and ${\cal{O}}(0,2)$. 

Initially one may wonder whether the CS correction can be gauged away via a coordinate 
transformation, though after some consideration it is clear this is not the case. 
In particular, one can calculate curvature invariants and show that these are indeed CS corrected 
to the order considered here. Perhaps the most obvious of these invariants is the Pontryagin density, 
$\pont$, which must be proportional to $\square \vartheta$, and so the deviation from Kerr
can readily be computed from (\ref{theta_11}). Also, as we shall see shortly, 
the location of the inner-most stable circular orbit is CS corrected, which is a further indication
that the CS modification is a non-trivial geometric perturbation of Kerr.

The location of the ergosphere and event horizon are however unchanged
by the CS correction to the order considered here. 
The ergosphere is the location where the Killing vector $(\partial/\partial t)^\alpha$
becomes null, or equivalently when $g_{tt}=0$, and this component of the metric
is unaltered by the CS-correction. The event horizon of the metric can be found by
tracing an outgoing quasi-spherical lightcone backwards in time from $\Scri^+$. Such a lightcone
is defined by an axisymmetric null hypersurface $u(t,r,\theta)\equiv t-R(r,\theta)={\rm const.}$
that satisfies $u_{,\alpha} u_{,\beta} g^{\alpha\beta}=0$ \cite{Pretorius:1998sf}. 
It is straight-forward to check that, to this order in the expansion parameters,
the lightcone equation does not depend on the CS correction, and thus will have the
same solutions as Kerr; hence the horizon location is unchanged.

An immediate consequence of the rapid $1/r^4$ decay of the perturbation
to the Kerr metric \eqref{sol:metric_elements} is that to this order the
asymptotic structure of the solution is unaltered, and thus, the CS corrected solution has the same
ADM mass and angular momentum~\cite{Arnowitt:1962}.
A subtle point, however, is that the ``physical'' mass and angular momentum need
not be given by the ADM quantities in alternative theories of gravity. In order to determine
the former, one needs to perform a study of the conserved Noether current and charges 
in the modified theory {\emph{a la}}~\cite{Tekin:2007rn}, which we shall not carry out here. 
Moreover, note that even though we colloquially called the scalar field part of the 
solution ``hair'', it does {\it not} constitute a violation of the no-hair
theorems in that we cannot freely choose $\vartheta$; to this order
in the perturbation $\vartheta$ is uniquely given once we choose $M$ and $a$.

Next, we compute the total energy carried by the scalar field, the
expression for which is
\be
E_{\vartheta} = \int_{\Sigma} T_{\alpha\beta}^{(\vartheta)} t^{\alpha} t^{\beta} \sqrt{\gamma} d^{3}x,
\ee
where $\Sigma$ is a $t=\rm{const.}$ hypersurface, $t^{\alpha}=(\partial/\partial t)^\alpha$, and
$\gamma$ is the determinant of the metric intrinsic to $\Sigma$.
Integrating the energy for the solution [Eqs.~\eqref{theta-sol-SR}-\eqref{sol:metric_elements}]
{\it outside} the horizon gives
\be
E_{\vartheta} = \frac{1685 \pi \kappa}{36864} \frac{a^{2}}{M^{2}} \frac{\xi}{M^{3}},
\ee
which is of ${\cal{O}}(2,1)$ and thus beyond what is considered here.
Thus, to this order in the expansion, the scalar field contributes zero total energy
to the spacetime, which is consistent with both the horizon area and ADM
mass not changing (and similar for the angular momentum), assuming the usual relationships 
between the geometric properties and mass/angular-momentum of the BH holds.
Note that if we require positive energy for $\vartheta$, the coupling constant $\xi$ must
likewise be positive.

We next turn to the geodesic structure of the spacetime.
One interesting modification 
is the location of the innermost stable circular orbit, or ISCO, for
equatorial orbits.
This can be calculated as the circular orbit with minimum energy
$E$. The pure GR result is given by~\cite{Bardeen:1972fi,Ori:2000zn}
\ba
\frac{R_{ISCO}}{M} &=& 3 + Z_{2} \mp \sqrt{(3 - Z_{1}) (3 + Z_{1} + 2 Z_{2})}
\nonumber\\
Z_{1} &=& 1 + \left(1 - a^{2} \right)^{1/3} \left[ \left(1 + a\right)^{1/3} + \left(1 - a\right)^{1/3}\right].
\nonumber \\
Z_{2} &=& \left(3 a^{2} + Z_{1}^{2} \right)^{1/2}.
\ea
To leading order in the expansion parameters, we find that in CS gravity the location of the ISCO is at
\be
R_{ISCO}=6M \mp \frac{4\sqrt{6}a}{3}
          -\frac{7a^2}{18 M}
         \pm \frac{77\sqrt{6}a\xi}{5184 M^4},
\ee
where the upper signs are for co-rotating geodesics, and the lower signs
for counter-rotating ones.
As discussed above, if we require that the CS scalar $\vartheta$ has a local
stress-energy tensor that satisfies the usual energy conditions, then $\xi$ must
be positive, and therefore the CS correction always {\it opposes} the change
in the location of the ISCO relative to the leading order change
that rotation introduces. I.e., for co-rotating geodesics, the 
CS correction has the effect of enlarging the ISCO, and vice-versa for
counter-rotating geodesics.
That the ISCO location is changed could offer potentially
observable consequences in accreting BH systems, as the inner properties
of the accretion disk are strongly affected by the location of the ISCO.

The new solution also modifies the
dragging of inertial frames by the rotation of the BH. This
can most easily be gauged by looking at the 
angular velocity $\omega_{Z}$ of zero-angular-momentum observers (ZAMOs);
$\omega_{Z} = -g_{t\phi}/g_{\phi\phi}$, or explicitly
to leading order in the expansion parameters:
\be
\label{ang-mom}
\omega_{Z} = \frac{2 M a}{r^{3}} - \frac{5}{8} \frac{a}{M} \frac{\xi}{M^{4}} \frac{M^{5}}{r^{6}} \left(1 + \frac{12}{7} \frac{M}{r} + \frac{27}{10} \frac{M^{2}}{r^{2}} \right).
\ee
As with the effect on the ISCO location, if $\xi$ is positive, 
the CS correction always {\it reduces} the magnitude of frame-dragging
relative to Kerr for geodesics that approach close to the horizon.

For a more quantitative exploration of the CS correction to the frame-dragging mechanism we
can apply the gravitomagnetic formalism.
Let us then first transform the metric to Cartesian coordinates via the standard transformation
\ba
x &=& r \left(1 + \frac{a^{2}}{2 r^{2}} \right) \cos{\phi} \sin{\theta},
\nonumber \\
y &=& r \left(1 + \frac{a^{2}}{2 r^{2}} \right) \sin{\phi} \sin{\theta},
\nonumber \\
z &=& r \cos{\theta},
\ea
expanded to leading order in ${\cal{O}}(a/M)$. We can now define the trace-reversed
metric perturbation
\be
\bar{h}_{ab} = h_{ab} - \frac{1}{2} \eta_{ab} h,
\ee
where $h = h^{a}{}_{a}$ is the trace of the metric perturbation, as well as the gravitomagnetic potential via $A_{a} = -\bar{h}_{a 0}/4$, 
where $0$ stands for the time-component. With this at hand, we can then define the gravitomagnetic field via
\be
B^{i} = \epsilon^{0 i j k} \partial_{j} A_{k},
\ee
which for the metric found here reduces to
\ba
B_{x} &=& \frac{3}{2} \frac{a M}{R^{2}} \frac{x\,z}{R^{3}} - \frac{15}{16} \frac{a \; x \; z \; \xi}{R^{8}},
\nonumber \\
B_{y} &=& \frac{3}{2} \frac{a M}{R^{2}} \frac{y\,z}{R^{3}} - \frac{15}{16} \frac{a \; y \; z \; \xi}{R^{8}},
\nonumber \\
\label{bz}
B_{z} &=& -\frac{1}{2} \frac{a M}{R^{3}} \left(1 - \frac{3 z^{2}}{R^{2}} \right) + \frac{5}{8} \frac{a \; \xi }{R^{6}} , 
\ea
where $R^{2} = x^{2} + y^{2} + z^{2}$. One can check that the GR terms are indeed the standard 
ones (see eg.~\cite{Yunes:2008ua}), 
being proportional to $J = M a$, while the CS correction is proportional to $J/M$, since it is mass-independent.
From this, we can see that the modification is comparable to a third post-Newtonian correction
to the frame-dragging effect, ie.~the CS correction is suppressed by a factor of $1/R^{3}$ relative to the GR 
correction, which for objects on circular orbits corresponds to a correction of ${\cal{O}}(v^{6})$, where $v$ is the 
circular orbital velocity. 

Again, the preceding calculation indicates that dynamical CS modified gravity is not subject
to meaningful bounds via weak-field tests. 
In order to explicitly verify this expectation, we can study some observable derived from the CS correction.
Here, following~\cite{Yunes:2008ua}, we calculate the variation of orbital elements of a binary, averaged over one period, by solving 
the Gaussian perturbation equations in the small eccentricity limit. Strictly speaking, these results can only
be applied in the limit of extreme mass ratios, with the massive object a slowly rotating BH.
We suspect that
the modified Kerr solution will also provide a good description to the dynamical-CS induced correction to the exterior field of a 
rotating compact matter object, such as a neutron star, though of course a solution for that scenario would need
to be found to verify the claim. 
We find that the relative average variation of any orbital element, $A$, can be decomposed into $A = A_{GR} + A_{CS}$, where the first term is the GR expectation
and the second term is the CS correction. We find then that only the average rate of change of the longitude of the line of nodes, $\dot{\Omega}$,  and the argument 
of the perigee, $\dot{\omega}$, are CS corrected, namely 
\be
\frac{\left< \dot{\Omega}_{CS}\right>}{\left< \dot{\Omega}_{GR}\right>} = \frac{\left< \dot{\omega}_{CS}\right>}{\left<\dot{\omega}_{GR}\right>} = \frac{25}{64} \frac{\xi}{M r_{sm}^{3}} \left(1 - \frac{9}{5} \cos^{2}\iota \right),
\ee
where $r_{sm}$ is the semi-major axis, $\iota$ is the inclination angle  and the angled-brackets stand for the average over one orbital cycle. 

It then follows that a measurement of $\iota$ and $\dot{\Omega}$ or $\dot{\omega}$ to an accuracy $\delta$ can be used to test CS modified gravity, 
as done for example in~\cite{Smith:2007jm,Yunes:2008ua}. Let us assume that the measurement is in full agreement with the GR expectation up to 
experimental uncertainties, and that $\iota = \pi/2$. Then, one can constrain the coupling strength of the CS correction to be
\be
\xi  \lesssim \frac{64}{25} M r_{sm}^{3} \; \delta.
\ee
We can go one step further and use the data derived from observations of PSR J0737-3039 A/B~\cite{Burgay:2003jj} to place the first bound on 
dynamical CS modified gravity, namely $r_{sm} = 4.24 10^{5} \; {\textrm{km}}$, $\iota \approx 86$ degrees and $M = 1.476 M_{\odot}$. We then find
\be
\label{constraint}
\xi^{1/4} \lesssim 10^{4} \; {\textrm{km}},
\ee
where we have used an extremely conservative measure of the cumulative error $\delta \sim 0.1$ degrees per year. 

Although this is the first bound on dynamical CS modified gravity, we wish to stress that there are several caveats that must be taken into account. First, as mentioned above, we have 
used a BH solution to represent the exterior gravitational field of a neutron star. 
Such an identification can be shown to be accurate for weakly-gravitating
systems in GR, however in CS modified gravity this need not be the case.
Second, in order to use binary
pulsar observations to test an alternative theory of gravity one would have to calculate at least two other observables, one of which is of course the rate of change
of the semi-major axis due to GW emission. The generation of gravitational radiation in CS modified gravity has not yet been worked out, and thus, this aspect 
cannot yet be incorporated into (\ref{constraint}).
In all then, the
constraint should be considered as a
{\it{conservative}} order-of-magnitude estimate on the the dynamical formulation of CS modified gravity.

\section{Conclusions}
\label{conclusions}

In this work we have presented a first study of the nature of slowly rotating BHs
in dynamical CS modified gravity, to leading order in the coupling constant.
In contrast to solutions in the non-dynamical framework, discussed in Appendix~\ref{rot-BH-non-dym},
no ad-hoc prescriptions need be made regarding the nature of the CS
field $\vartheta$. Instead, we only demand the physically reasonable
requirements that $\vartheta$ respect the symmetries
of the spacetime, and posses finite, positive energy exterior to the BH
event horizon. The resultant solution represents an inherently {\it strong-field}
perturbation of Kerr, in that the deformation of the background geometry
decays as $(M/r)^4$. As such, the deformation can easily be consistent with all existing weak-field
bounds, yet still offer the exciting possibility of allowing very different
phenomena in strong-field scenarios involving spinning BHs:
compact object mergers, the inner edges of accretion disks, gravitational
collapse, etc. 

Regarding the nature of the solution, we found that to the order of the 
perturbation studied here, the horizon and ergosphere locations are unaltered relative to Kerr,
and similarly the ADM mass and angular momentum
of the spacetime is unchanged. The dynamical field $\vartheta$ can be thought-of
as imbuing the BH with dipole scalar ``hair'', though the structure of
which is uniquely determined by
the spin and angular momentum of the BH. 
The main observable effect
of the CS correction is to alter the near-horizon geodesic structure of the spacetime,
effectively weakening the frame-dragging phenomenon, and moving the ISCO farther out (closer in) 
for co-rotating (counter-rotating) orbits relative to Kerr.

We further calculated the effect of the CS correction on the evolution of the parameters
describing a binary orbit relative to GR. We then used these results with binary pulsar observations
to place a bound on the strength of the coupling constant in the dynamical
theory of $\xi^{1/4} \lesssim 10^{4} \; {\textrm{km}}$. The caveats with this application of the new solution are that
we have not yet studied whether it also gives an adequate description
of the CS correction to the geometry exterior to a neutron star, that the orbit can
be depicted as a geodesic about a slow-rotating BH, and that 
to this order the CS modification to GW emission in the system is
negligible.

Future work includes continued study of the nature of the dynamical theory, in particular
BH solutions and compact object mergers, without imposing any slow-rotation or 
small-coupling limit approximations. This will likely require numerical solution
of the CS modified Einstein equations. One important question that such studies
could shed light on is how does the weak-field nature of a well-defined metric
theory of gravity restrict the nature of the strong field regime for
astrophysically relevant solutions. 

Such a study is of particular importance to the budding
field of GW astronomy. On the one hand GW detectors, such as LIGO (Laser Interferometer Gravitational Observatory)~\cite{Abramovici:1992ah,Waldmann:2006bm,:2007kva}
or the planned space-based detector LISA (Laser Interferometer
Space Antenna)~\cite{Bender:1998,Danzmann:2003tv,Prince:2003aa} could be
used to place restrictions on potential CS deviations from GR. On the other hand, such studies could also help us understand
how seriously GW detection might be negatively impacted by search strategies
built around the assumption that GR is the correct theory of classical space and time;
i.e.~, could classes of strong-field GW sources be misidentified,
or even missed altogether if there are significant deviations from GR in the strong-field regime?

Investigations of this type could also help clarify how robustly conclusions can be made about
the nature of strong-field sources. For example, let us assume that we know BH mergers
``look'' similar in terms of GW emission for a large class of metric
theories that only differ from GR in the strong-field regime. Then, 
following detection
of such a source, one can have significant confidence in a claim that {\it black holes}
have been directly observed, even if one cannot claim they are Kerr black holes.

{\bf{Note added in revision:}} After this paper was submitted, another 
work~\cite{Konno:2009kg} appeared on the arXiv that independently derived
the solution presented in Sec.~\ref{slow-rot}.

\acknowledgments

We are grateful to Paul Steinhardt, Vitor Cardoso, Paolo Pani, Carlos Sopuerta, Shaun Wood and Bogdan Stoica for useful suggestions and comments.  Some calculations used the computer algebra systems MAPLE, in combination with the GRTensorII package~\cite{grtensor}. We acknowledge support of the NSF grant PHY-0745779, and FP acknowledges support from the Alfred P. Sloan Foundation.

\appendix
\section{Rotating BH Solutions in Non-Dynamical CS Modified Gravity }
\label{rot-BH-non-dym}

In this appendix, we explore and generalize previous results regarding rotating BH solutions in the non-dynamical framework. First, we review the slow-rotating solution of KMT and extend it to much more general CS scalar fields. Second, we find a new solution for the metric tensor with the KMT choice of CS scalar, thus proving that the KMT solution is not unique. We also show that the solutions
found in the non-dynamical framework do not remain solutions in the dynamical theory.

In the non-dynamical framework, KMT used the slow-rotation approximation to find a stationary and axisymmetric solution, provided the CS coupling field was chosen to be non-canonical. In particular, the KMT choice of $\vartheta$ is
\be
\label{KMT-th}
\vartheta^{\textrm{KMT}} = \frac{r \cos{\theta}}{\lambda_{0}},
\ee
where $\lambda_{0}$ is a constant. This choice of CS scalar leads to a solution where all metric perturbations vanish, except for the $g_{t\phi} = - r^{2} \bar{\omega}$ components, where 
\be
\label{K-omega}
\bar{\omega}^{KMT}(r) = \frac{B_{2}}{r^{2}} f 
+ \frac{B_{1}}{r^{3}} \left[ r^{2} f - 4M^{2} + 4 M r f \ln(r f)\right],
\ee
and $B_{i}$ are integration constants. 

This solution was later used to propose an explanation to the flat rotation curves of galaxies~\cite{Konno:2008np} as follows. 
Consider the transverse, circular velocity of point particles
in this background $v_{\phi}$, in the geodesic approximation and neglecting $\vartheta$-backreaction. In the far-field limit, $M/r \ll1$, and using the metric found above~\cite{Konno:2008np},
\be
\label{vphi-KMT}
v_{\phi} = \sqrt{\frac{M}{r}} + r \bar{\omega} + \frac{r^{2}}{2} \bar{\omega}_{,r}  \sim \sqrt{\frac{M}{r}} + {\textrm{const.}},
\ee
where the first term is the Schwarzschild term and the second term is a CS correction, proposed as an explanation to the galactic rotation curves. 

The KMT solution is an interesting result that sheds some light on the effects of CS modified gravity on certain observables, though before one can claim victory over the rotation curves, their analysis must be considered more carefully. In doing so, we have found that the KMT solution suffers from a few drawbacks that render it unphysical as a true BH solution. The main problem stems from the fact that this solution was found initially in the non-dynamical formulation, which as we argued in the introduction is quite contrived, arbitrary and probably not well-posed. The hope would then be to embed this solution in the dynamical framework, however such a task is impossible as we explain below. 

In the dynamical formulation the nature of the CS field $\vartheta$ changes in the following fundamental way:
$\vartheta$ becomes a dynamical field governed by a scalar field Lagrangian, and consequently 
provides a new stress-energy contribution to the CS modified Einstein equations.
The KMT solution indeed satisfies the evolution equation for $\vartheta$ to leading order in the spin~\cite{Konno:2008np}. The stress-energy tensor associated with $\vartheta_{KMT}$, however, has infinite total energy, and therefore the $KMT$ metric
is not a self-consistent solution to the dynamical field equations.
 
In view of these problems, one could wish to generalize the slow-rotation results in~\cite{Konno:2007ze}, by considering more general CS scalars. 
A straightforward calculation reveals that the most general $\vartheta$ for which the solution found in~\cite{Konno:2007ze} persists is 
\be
\label{gen-vartheta}
\vartheta^{\textrm{gen}} = A_{0} + A_{x} r \cos{\phi} \sin{\theta} + A_{y} r \sin{\phi}  \sin{\theta}  + A_{z} r  \cos{\theta},
\ee
where $A_{i}$ are constants. In fact, we can rewrite this CS coupling field as $\vartheta = \delta_{ab} A^{a} x^{b}$, where $x^{a} = [1,x,y,z]$ and
$\delta_{ab}$ is the Euclidean metric. 
Obviously, although more general, this scalar contains the same problems as $\vartheta_{KMT}$, namely its associated stress-energy leads to 
infinite energy, making it incompatible with the dynamical framework. 

Additionally, one can also show that there is another solution to the modified field equations that was missed in~\cite{Konno:2007ze}. 
In particular, a direct calculation reveals that 
\ba
\vartheta &=& \bar{f}(r,\phi) + r \bar{g}(\phi) + r \bar{h}\left(C_{1} \phi - t\right) +  r \bar{k}(\theta,\phi)
\nonumber \\
&+& r \int \frac{dr}{r} \left[ -\partial_{r} \bar{f}(r,\phi)+ \frac{1}{r} \bar{f}(r,\phi) + \frac{1}{r} \bar{j}(r)\right] , 
\nonumber \\
\label{w-good}
\bar{\omega} &=& -\frac{C_{1}}{r^{2}} f,
\ea
also solves the modified field equations, where $\bar{f}$, $\bar{g}$, $\bar{h}$, $\bar{j}$ and $\bar{k}$ are arbitrary functions and $C_{1}$ is another integration constant. 
The CS coupling field found here does bypass the problem with the stress-energy tensor, since the arbitrary functions can be freely chosen 
to decay fast enough. For example, if $\bar{f} = \bar{g} = \bar{h} = \bar{k} = 0$ and $\bar{j}=-3 j_{0}/r^{2}$, then $\vartheta = j_{0}/r^{2}$, for some constant $j_{0}$. Thus, although the canonical $\vartheta$ is not allowed by this family of solutions, Eq.~\eqref{w-good} is compatible with the
dynamical framework. 

This better-behaved and much more general CS coupling, however, leads to a solution that differs sufficiently from the KMT one to
lose its appeal. In other words,  the new solution in Eq.~\eqref{w-good} cannot explain the galactic rotation curves because the 
logarithmic term has disappeared from the metric. One can show by direct calculation that
\be
v_{\phi} \sim \sqrt{\frac{M}{r}} - C_{1}  \frac{M}{r^{2}},
\ee
in the far-field limit $M/r \ll 1$. This result is to be contrasted with Eq.~\eqref{vphi-KMT}, which leads to a constant term instead of a term
that decays quadratically with radius.  Due to this decaying behavior, the new solution cannot explain the flat rotation curves.

One could argue that the above analysis is evidence for nature selecting the KMT solution, instead of the one presented here, 
but we wish to argue precisely the opposite. In the non-dynamical framework, there is nothing to suggest that the 
KMT solution is more valid or less valid than the new solution. The fact that two different observables are obtained is due to the immense freedom
present in the choice of CS scalar, which points at an incompleteness of the non-dynamical framework. This is to be contrasted with the 
dynamical framework, where the coupling field is determined by its evolution equation and there is only freedom in specifying
its initial condition.

\section{Arbitrarily Fast-Rotating BH Solutions}
\label{rot-fast}

In this appendix we study the possibility of extending the analysis presented in this paper to a background that is not necessarily 
slowly-rotating. We shall see, however, that dropping the slow-rotation approximation is sufficient to prevent us from being able 
to find analytic solutions. 

Consider then the Kerr metric in Boyer-Lindquist coordinates:
\ba
\label{kerr-pert}
ds^{2} &=&  - \left(1 - \frac{2 M r}{\Sigma} \right) dt^{2} - \frac{4 M \, a \, r \sin^{2}{\theta}}{\Sigma} dt d\phi + \frac{\Sigma}{\Delta} dr^{2} 
\nonumber \\
&+& \Sigma d\theta^{2} + \sin^{2}{\theta} \left(r^{2} + a^{2} + \frac{2 M r a^{2} \sin^{2}{\theta}}{\Sigma} \right) d\phi^{2},
\nonumber \\
\ea
where $\Sigma = r^{2} + a^{2} \cos^{2}{\theta}$ and $\Delta = r^{2} - 2 M r + a^{2}$. The perturbation shall be parameterized in 
the same way as before, leading to the following full metric:
\ba
ds^{2} &=&  - \left(1 - \frac{2 M r}{\Sigma} \right) \left(1 + h \right) dt^{2} 
\nonumber \\
&-& \frac{4 M \, a \, r \sin^{2}{\theta}}{\Sigma} \left(1 + \omega \right) dt d\phi 
\nonumber \\
&+& \frac{\Sigma}{\Delta} \left(1 + m \right) dr^{2} + \Sigma \left(1 + k \right) d\theta^{2} 
 \\ \nonumber
&+& \sin^{2}{\theta} \left(r^{2} + a^{2} + \frac{2 M r a^{2} \sin^{2}{\theta}}{\Sigma} \right) \left(1 + p \right) d\phi^{2}.
\ea
All arbitrary metric perturbations, $h$, $\omega$, $m$, $k$ and $p$, are assumed to be at least of ${\cal{O}}(\zeta)$.

We can now study the evolution equation for the CS coupling field in the small-coupling approximation. In this limit, the evolution equation
becomes
\ba
\Delta \; \vartheta_{,rr} &+&  2  \left( r - M \right) \vartheta_{,r} + \cot{\theta} \; \vartheta_{,\theta} +  \vartheta_{,\theta\theta}
\nonumber \\
&=& - \frac{24 \alpha}{\beta} \left(a \, M^{2} \right) \left(r \cos{\theta} \right) 
\nonumber \\
&\times& 
\frac{\left(r^{2} - 3 a^{2} \cos^{2}{\theta} \right) \left(3 r^{2} - a^{2} \cos^{2}{\theta} \right)}{\Sigma^{5}}.
\ea
The general solution to any such inhomogeneous differential equation will be given by the linear combination of a homogeneous and a particular solution. The homogeneous solution is 
\be
\vartheta_{H} = \tilde{\vartheta}_{H}(r) \hat{\vartheta}_{H}(\theta),
\ee
where
\ba
\tilde\vartheta_{H}(r) &=& E_{1} F\left[\left[\frac{\tilde\alpha}{2},\frac{\tilde\alpha}{2}\right],\tilde\alpha,\chi \right] \left(r - M + \gamma\right)^{-\tilde\alpha/2} 
\nonumber \\
&+& E_{2} F\left[\left[\frac{\tilde\beta}{2},\frac{\tilde\beta}{2}\right],\tilde\beta,\chi \right] \left(r - M + \gamma\right)^{-\tilde\beta/2},
\nonumber \\
\hat\vartheta(\theta) &=& E_{3} P_{-\tilde\alpha/2}(\cos{\theta}) + E_{4} Q_{-\tilde\alpha/2}(\cos{\theta}), 
\ea
where again $P(\cdot)$ are Legendre polynomials of the first kind, $Q(\cdot)$ are Legendre polynomials of the second kind, 
$F(\cdot)$ are generalized hypergeometric functions, $E_{i}$ are constants of integration and the coefficients $\tilde{\alpha}$ and 
$\tilde{\beta}$ are given in Eq.~\eqref{tilde-alpha}. We recognize this solution as a generalization of Eq.~\eqref{Hom-sol-1} to 
arbitrarily rotating BHs, where 
\ba
\gamma &=& \sqrt{M^{2} - a^{2}},
\nonumber \\
\chi &=& \frac{2 \gamma}{r - M + \gamma}.
\ea
Due to the strong similarities between this solution and the slow-rotation one, we directly conclude that the constants
of integration must be chosen such that $\vartheta_{H} = {\textrm{const.}}$, by requiring asymptotic flatness, reality and a 
well-defined stress-energy tensor. 

The problem then reduces to finding the particular solution to the evolution equation for arbitrarily fast BH rotation. 
Unfortunately, such an equation does not appear to be amenable to symbolic manipulation. 
Without such an analytic solution, we cannot proceed with the boot-strapping algorithm described in the main paper.
However, the differential operator associated with the Ricci and C-tensors for the metric in Eq.~\eqref{kerr-pert} is truly formidable. Thus, the probability 
of finding a solution for the metric perturbation, even given an analytic form for $\vartheta$, is small. 


\bibliographystyle{apsrev}
\bibliography{review}

\end{document}